# Light induced decoupling of electronic and magnetic properties in manganites

H. Navarro[1*], Ali C. Basaran[1], F. Ajejas[1], L. Fratino[2,3], S. Bag[2], T. D. Wang[1], E. Qiu[1], V. Rouco[4], I. Tenreiro[4], F. Torres[5,6], A. Rivera-Calzada[4], J. Santamaria[4], M. Rozenberg[2] and Ivan K. Schuller[1]

[1] Department of Physics, Center for Advanced Nanoscience, University of California, San Diego, 92093, USA.

[2] Université Paris-Saclay, Centre National de la Recherche Scientifique Laboratoire de Physique des Solides, 91405 Orsay, France.

[3] Laboratoire de Physique Théorique et Modélisation, CNRS UMR 8089, CY Cergy Paris Université, 95302 Cergy-Pontoise Cedex, France.

[4] Departamento de Física de Materiales, Universidad Complutense de Madrid, 28040 Madrid, Spain.

[5] Department of Physics, Universidad de Chile, Santiago 7800024, Chile.

[6] Center for the Development of Nanoscience and Nanotechnology, CEDENNA, Santiago 9170124, Chile.

Corresponding Author

*E-mail: hnavarro@physics.ucsd.edu

The strongly correlated material $La_{0.7}Sr_{0.3}MnO_3$ (LSMO) exhibits metal-to-insulator and magnetic transition near room temperature. Although the physical properties of LSMO can be manipulated by strain, chemical doping, temperature, or magnetic field, they often require large external stimuli. To include additional flexibility and tunability, we developed a hybrid optoelectronic heterostructure that uses photocarrier injection from cadmium sulfide (CdS) to an LSMO layer to change its electrical conductivity. LSMO exhibits no significant optical response, however, the CdS/LSMO heterostructures show an enhanced conductivity, with ~ 37 % resistance drop, at the transition temperature under light stimuli. This enhanced conductivity in response to light is comparable to the effect of a 9 T magnetic field in pure LSMO. Surprisingly, the optical and magnetic responses of CdS/LSMO heterostructures are decoupled and exhibit different effects when both stimuli are applied. This unexpected behavior shows that heterostructuring strongly correlated oxides may require a new understanding of the coupling of physical properties across the transitions and provide the means to implement new functionalities.

## 1. Introduction

Strongly correlated electron systems display various physical phenomena such as superconductivity, ferroelectricity, ferromagnetism, and antiferromagnetism as a consequence of the intertwining of several degrees of freedom (charge, spin, strain, doping), producing a wide variety of phases and resulting in complex phase diagrams[1-6]. The critical response of these materials to a modification of the parameters makes them promising candidates for the next generation of electronic devices[7-12]. However, many of these variables are determined by intrinsic material properties set during growth and cannot be easily controlled in applications. Additional functionalities can be introduced by post-growth modifications[8, 13] or by interfacing the correlated oxides with other functional materials with very high sensitivity to external stimuli, such as light[12, 14-16].

Manganites are one of the paradigmatic examples of strongly correlated electronic materials, showing a characteristic magnetoresistive behavior[17, 18]. In particular, $La_{0.7}Sr_{0.3}MnO_3$ (LSMO) is one of the most studied materials for spintronic applications[19-21] due to its unique ferromagnetic and half-metallic order up to 360 K. The physical properties of LSMO can be tuned by modifying the electrical, magnetic, strain, and lattice degrees of freedom[22-27]. While the magnetic and electronic properties of LSMO are correlated, in practice, it is often inefficient to control one with the other. In addition, spatial and temporal control of temperature with a certain precision often requires complex architectures and is not energy efficient[9],[28]. Although LSMO is not sensitive to light, some oxygen-deficient LSMO thin films, exhibit a persistent increase of the electrical conductivity after light illumination at low temperature[23]. Moreover, LSMO/$MAPbI_3$ heterostructures have been change their magnetization with light illumination[22]. These results showed a great potential of hybrid heterostructures enabling optical functionalities with LSMO in technological applications for advanced oxide electronics[12, 14]. This motivates the search for new functionalities by using light to control the electrical and magnetic properties of strongly correlated materials.

Recently, we reported a large modulation of the MIT temperature ($T_c$), for both non-volatile and volatile resistive switching, of Mott-insulators $VO_2$, $V_2O_3$[15], and $V_3O_5$[29] by incorporating them into a heterostructure with photoconductive CdS[30]. Thus, motivated by the expectation of emergent proximity-induced interfacial effects, we heterostructured LSMO with light-sensitive CdS. We found that the resistance in the insulating state drops significantly when these heterostructures are illuminated with visible light while the transition temperature ($T_c$) remains constant. The volatile change in the resistance upon illumination depends on the LSMO thickness, suggesting that this is an interface effect. On the other hand, contrary to the effect of light while a magnetic field lowers the resistance it also shifts the transition temperature to higher temperatures. We propose a model where the density of charge carriers increases upon light illumination, resulting in enhanced conductivity. Our results show a practical way to control the electronic properties of manganites with light while also suggesting a new way to understand the coupling between electronic and magnetic properties in these strongly correlated systems.

## 2. Results

We measured the resistance versus temperature (R(T)) in multiple CdS/LSMO samples with different LSMO thicknesses as shown in the inset of Figure 1c). The results of the hybrid sample with 3.5 nm LSMO thickness are plotted in Figure 1. Figure 1a) shows the R(T) of the heterostructure when in the dark and illuminated with visible light of different powers. For all measurements, the resistance as a function of temperature is non-hysteretic, as recorded for both warming and cooling branches. In the dark (green curve), the hybrid CdS/LSMO exhibits a metal-to-insulator transition at 280 K. This behavior is expected from pure LSMO[31, 32] and is in good agreement with the measured results of the control sample (pure LSMO thin film) shown in Supplementary Material. Such agreement confirms that the top CdS layer does not modify the LSMO electronic properties. With increasing light power density, the resistance of the insulating ground state decreases while the $T_c$ (defined as the maximum in the resistivity vs T) remains constant (Figure 1a)). The resistance drops more than 37 % at $T_c$ when the CdS/LSMO is exposed to a white LED light with 731 mW/cm$^2$ illumination power density. In the absence of CdS, the effect of light on pure LSMO is negligible (Figure S1a), which also indicates that heating effects are not significant.

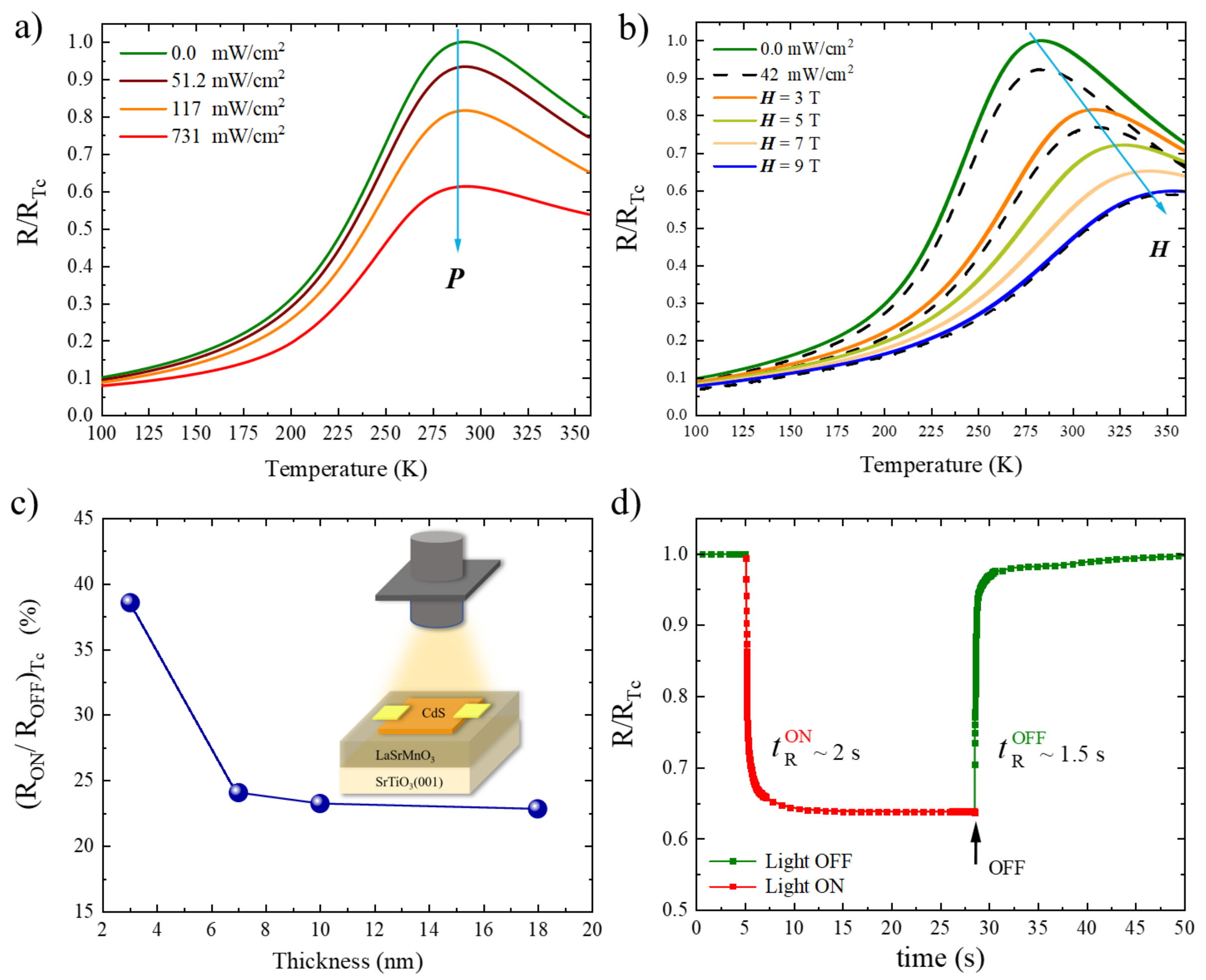

Figure 1: Light-induced modification of the metal-to-insulator transition in CdS/LSMO heterostructures. Resistance vs. Temperature R(T) measurements for hybrid CdS/LSMO heterostructures, a) varying light power density and b) varying magnetic field. The dashed lines in b) shows the effect of 42 mW/cm$^2$ light for selected magnetic fields (0. 3, and 9 Tesla). The solid green curves correspond to resistance vs. temperature without light in all panels. c) Change in the resistance at $T_c$ = 280 K, with illumination for different LSMO thicknesses. d) Relaxation of the resistance (normalized to resistance at $T_c$ = 280 K).

Figure 1b) shows the resistance versus temperature of the CdS/LSMO heterostructure (with 3.5 nm thick LSMO) when an external magnetic field is applied (3 to 9 T). A larger applied magnetic field results in a more significant resistance drop. For instance, for $\boldsymbol{H}$ = 3 T, the resistance at $T_c$ decreases by 20 %, shifting the $T_c$ of 23 K towards higher temperatures. This effect increases as the applied field increases, reaching a 40 % resistance drop at $T_c$ when a 9 T field is applied. Changes in $T_c$ to higher temperatures are present for all applied fields. In Figure 1b) all solid lines are acquired in the dark. Interestingly, when the sample is subjected to both magnetic field and light (sequentially), the responses to the two stimuli is additive. The sample exhibits a ~ 20 K shift in $T_c$, similar to the dark, but an additional 5 % resistance drop still appears under illumination (represented by the dashed curves). However, the sample would cease to respond to light if a 9 T field is applied previously. Such behavior is shown in Figure 1b).

The percentage decrease of resistivity with light depends strongly on sample thickness ~ 25 % at 7 nm, ~ 24 % at 10 nm, and 23 % at 18 nm (Figure 1c)). The absolute changes are shown in the Supplementary Material. The decrease in the resistance change under the same illumination with increasing LSMO thickness indicates that this is an interfacial effect, which will be discussed later.

The relaxation time dependence of the CdS/LSMO resistance after light exposure is shown in Figure 1d). The change in the resistance before (green curve) and during illumination (red curve) at 280 K shows a decrease of 40 % in less than 2 s. Once the minimum resistance is reached, the optical light source is turned off, recovering its original resistance in ~1.5 s. This shows that the effect of the light in the CdS/LSMO heterostructure resistance is volatile.

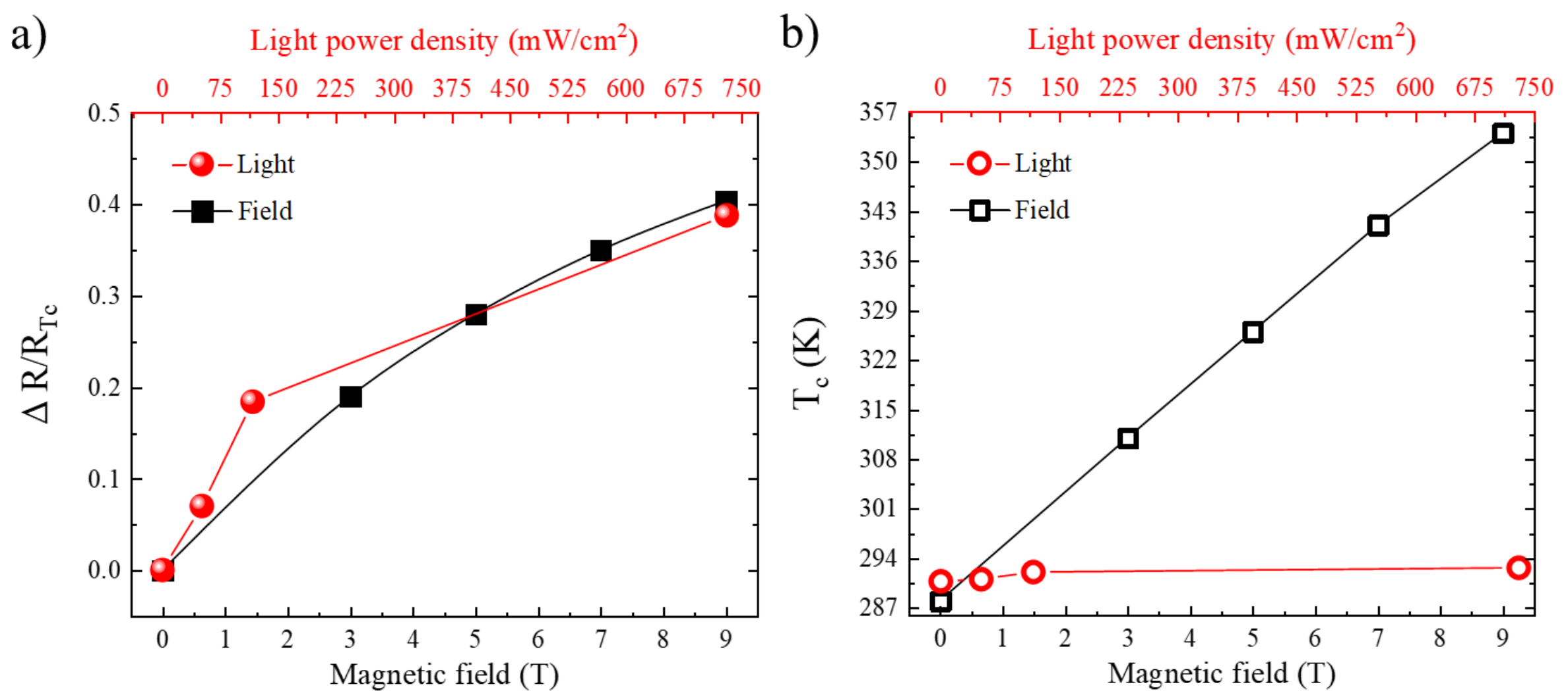

Figure 2: Comparison between two external stimuli in hybrid CdS/LSMO heterostructure. a) Normalized resistance changes $\Delta R$ at $T_c$ as a function of applied magnetic field (black squares) and light power density (red circles). b) $T_c$ as a function of applied magnetic field (black open squares) and power density (red open circles).

Figure 2 compares the effect of the two external stimuli, e.g., light and magnetic field, on the hybrid CdS/LSMO heterostructures. Figure 2a) shows the normalized resistance drop $\Delta R/R_{Tc}$ at the transition temperature under illumination (solid red circles) and applied magnetic field (black squares). For both external stimuli cases, the resistance decreases with increasing amplitude. 731 mW/cm$^2$ light power density decreases the resistance by 40 % at $T_c$, nearly the same as applying a very large 9 T external magnetic field.

To further compare the response of electrical properties to the two stimuli independently, Figure 2b) shows the dependence of $T_c$ on each stimulus. Light illumination had negligible impact on $T_c$ with a shift of 3 K at the maximum applied power density $P$ = 731 mW/cm$^2$. In contrast, the transition temperature increases linearly with the applied magnetic field. The $T_c$ shift can be as large as ∼ 67 K when a 9 T external field is applied. This surprising result confirms the different origins of the two mechanisms.

The resistance decrease is expected to be driven by carrier injection from CdS to LSMO film. Similar to our previous report on CdS/$VO_x$[29], the light response observed in CdS/LSMO is an interfacial effect between the CdS and LSMO layers. We observed that the response to light becomes smaller with LSMO layer thickness beyond 18 nm. However, in contrast to the present work, in the CdS/$VO_x$ case, light stimulation changes the resistance and the transition temperature together. This qualitative difference calls for a different physical model to explain the results found for CdS/LSMO.

Furthermore, when we subjected our sample to light while applying an external field (see Fig 1(b)), the effect of light appears in the same way as without a field, namely, an increase of conductivity throughout the whole temperature range without a shift in transition temperature. However, when a larger field is applied, the light response becomes less pronounced. Below, we shall describe a model calculation to understand the cause of this behavior.

**3. Discussion**

Generally, the electronic properties of $La_{1-x}Sr_xMnO_3$ may be described by the Double Exchange Model[33, 34]. Because of the Sr doping, the Mn atoms are in a mixed 3+/4+ valence state. Thus, three electrons of the 3*d t*2*g* band form a "localized" spin 3/2 state at each site. On the other hand, there are 1-*x* electrons per site occupying the conduction $e_g$ band, therefore, there are nominally *x* doped conduction hole carriers. The reason for this is the strong Hund's rule coupling. Thus, the conduction holes can be either strongly scattered or not by the localized spins 3/2, depending on whether their spins are aligned antiparallel or parallel, respectively.

Hence, when the core 3/2 spins are all aligned into a ferromagnetic state below $T_c$, the conduction electrons align to the direction of their magnetic moment $m_z$. In contrast, above the transition temperature, the core electrons are in a disordered paramagnetic (PM) state, and the conduction electrons suffer continuous scattering from the unavoidable misalignment of the spin moments. This explains qualitatively the MIT that characterizes the CMR manganites as they cross $T_c$.

We have incorporated these qualitative features in a phenomenological model to describe the results in the CdS/LSMO heterostructures. The density of conduction carriers is given by the doping $x$, which in the heterostructures is 0.3 per Mn site. Since the photoconductive CdS increases its number of free carriers upon illumination, we assume that part of the photoexcited carriers is transferred into the manganite thin film by a proximity effect. Thus, from the simple Drude model we have,

$$\sigma = \frac{ne^2\tau}{m}, \qquad (1)$$

where $n$ is the carrier density, $\tau$ is average time between ionic collisions. Further assuming that the additional carrier density from the CdS, $n_0\Gamma$, is simply proportional to the illumination power, we have,

$$n = n_0\,(1 + \Gamma), \qquad (2)$$

and therefore, the resistivity is modulated by the effects of the light as:

$$\rho_\Gamma = \rho_0 \,/\, (1 + \Gamma), \qquad (3)$$

To take into account the connection between magnetic behavior and resistivity, we adopt a slightly simplified version of the expression introduced by Wang et al[35], which was motivated by the analysis of experimental data in terms of the Double Exchange Model[33, 34]:

$$\rho_0 = ATe^{\frac{\epsilon(1-m_z^2)}{T}}, \qquad (4)$$

where $\epsilon$ is the activation energy, $T$ is the temperature, and $A$ is a fitting parameter. The physical content of this formula is clear. At high $m_Z$ ~1, deep in the FM phase, the exponent of this formula is small and a first order expansion of the exponential, recovers the semi-empirical form of the resistivity[31]. On the other hand, at small $m_Z$ the resistivity crosses-over to the activated form of a semiconductor.

The magnetization is governed by a ferromagnetic–paramagnetic transition at $T_c$. The effective coupling $J$ between the local spins of Mn is related to the Hund's coupling $J_H$ and the carrier number[31]. Since in LSMO at $x \sim 0.3$ the $T_c$ depends weakly on $x$, we can take the parameter $J$ as a constant[36]. Then, for simplicity, we model the temperature- and external field $h$-dependent FM transition by means of an Ising model solved in mean-field approximation:

$$m_z = tanh\left(\frac{h + 4Jm_z}{T}\right), \tag{5}$$

which gives for the critical temperature $T_c = 4J$ at $h = 0$. We may now adopt suitable parameters for this model and explore its predictions.

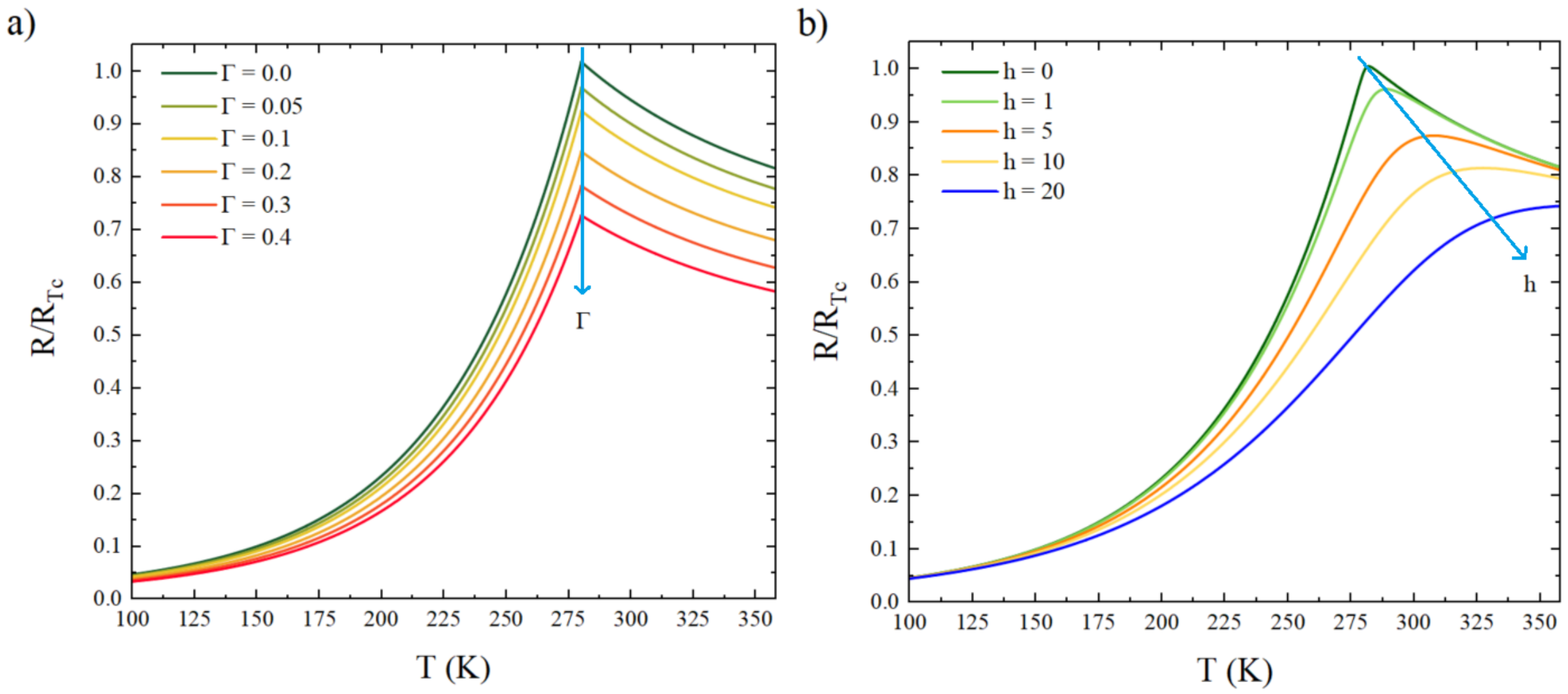

Figure 3: Theoretically computed metal-insulator transition. a) Change in the resistivity of LSMO as a function of temperature for different carrier density Γ, b) and for different applied fields, h.

Figure 3a) shows the results from our model, which shows a decrease in the resistivity proportional to the applied light power without a shift in $T_c$, capturing the qualitative features of the experiment. As a further test of our model, we checked the effects of an external magnetic field h, as shown in Figure 3b). The magnetic field shifts the $T_c$ to higher temperatures, also in good qualitative agreement with the experiments. The reason is that the external field provides further stability to the ferromagnetic phase. Thus, $m_z$ survives at higher temperatures (see

Supplemental Material Figure S5 for magnetization measurement as a function of thickness). Consequently, the maximum of the $\rho$(T) that defines the MIT is shifted to higher temperatures.

## 4. Conclusion

In conclusion, we demonstrated that the electronic conductivity of a hybrid optoelectronic heterostructure CdS/LSMO is modified by photocarrier injection from the CdS into the LSMO layer when exposed to light. Although the magnetic and electrical properties in pure LSMO are believed to be correlated, we can independently control the electrical and magnetic properties. While LSMO is known to have no significant optical response, our ultrathin film CdS/LSMO heterostructures show increased light induced conductivity, with ~37 % resistance drop at the transition temperature. The increase in conductivity with light is volatile and equivalent to applying a 9 T magnetic field in this colossal magnetoresistance material. Under a magnetic field, the transition moves to higher temperatures because of the increased stability of the ferromagnetic phase. The changes in the resistance under optical and/or magnetic stimuli are surprisingly independent of each other. Moreover, as the application of large magnetic fields is not readily achievable in most applications, our finding demonstrates a new practical way to control electronic properties with light in complex oxides.

## 5. Experimental

The ultrathin epitaxial LSMO samples (3.5, 7, 10, 18 nm) were grown on STO ⟨100⟩ substrates using a high-pressure (3.2 mbar pure oxygen) and high-temperature (750°C) sputtering. The target-to-substrate distance was set to 1.5 cm, so the highly confined oxygen plasma was tangential to the substrate. 80 nm thick CdS films were grown in a separate deposition with rf magnetron sputtering from a CdS target in a 2-mTorr pure argon atmosphere at 150 °C. For each CdS/LSMO bilayer, two Au (40 nm) electrodes were patterned on the CdS/LSMO heterostructured films.

Morphological characterization measurements were performed in a Veeco Scanning Probe Microscopy (SPM) by analyzing the topographic information in LSMO samples and CdS/LSMO heterostructures. The SPM images can be found in Supplemental Material at Figure S3 for tapping mode. The structure of the films was analyzed by XRD in a Rigaku SmartLab system at room temperature (See Supplemental Material at Figure S4 for structural characterization). Transport measurements were carried out on a Montana C2 S50 Cryocooler and TTPX Lakeshore cryogenic probe station, using a Keithley 6221 current source and a Keithley 2182A nanovoltmeter (see Supplemental Material Figure S1 and S2 for electrical transport measurements). A Thorlabs white LED was used for the photoconductivity measurements with a wavelength range from 400 to 700 nm. The magnetic characterizations and resistance measurements with applied field were performed in a Quantum Design DynaCool system equipped with a 9 T superconducting magnet, an optical probe, and a vibrating sample magnetometer.

**Acknowledgments**

The synthesis of the heterostructures, measurement of transport, magnetic and optical properties were supported by the U.S. Department of Energy's Office of Basic Energy Science, under Grant DE-FG02-87ER45332, the synthesis of LSMO by the Spanish MCI through Grant nos. MAT 2017-87134-C02, PCI 2020-112093 and theory by the French ANR "MoMA" project ANR-19-CE30-0020, Chilean FONDECYT - 1211902 and Basal AFB220001.

**Author contributions**

HN and IKS conceived the idea. HN designed the experiment. HN, IT, VR and AR-C fabricated the samples. HN, EQ, FA, ACB, TW, performed the transport and magnetic measurements. The X-rays diffraction and SPM measurements were performed by HN, EQ. Simulations and theoretical aspects were performed by LF, SB, FT and MR. HN and IKS wrote the manuscript. All authors participated in the discussion of the results and corrected multiple iterations of the manuscript.

**Competing interests**

The authors declare no competing interests.

Light induced decoupling of electronic and magnetic properties in manganites

Supplemental Materials

H. Navarro[1*], Ali C. Basaran[1], F. Ajejas[1], L. Fratino[2,3], S. Bag[2], T. D. Wang[1], E. Qiu[1], V. Rouco[4], I. Tenreiro[4], F. Torres[5,6], A. Rivera-Calzada[4], J. Santamaria[4], M. Rozenberg[2] and Ivan K. Schuller[1]

Figure S1 shows the electrical transport as a function of the temperature for a 3.5 nm pure LSMO film with or without light illumination (Figure S1a)). In contrast to when a photoconductive material (CdS) is deposited on top of the LSMO, the pure LSMO film does not show any resistance change in metallic or insulating state when exposed to light. Hybrid CdS/LSMO heterostructures exhibit resistance changes in the insulating metal state in all LSMO samples (3.5, 7, 10, 18 nm) (Figure S1b)). The thickest sample with 18 nm LSMO shows less than a 25% resistance drop in the insulating state when exposed to light. On the other hand, the CdS/LSMO hybrid with 3.5 nm LSMO-shows the most significant change in resistance at room temperature at a maximum light intensity, which is above 37 % when illuminated. Figure S2 shows the electrical transport as a function of the temperature of the pure CdS with and without light confirming its semiconductor behavior.

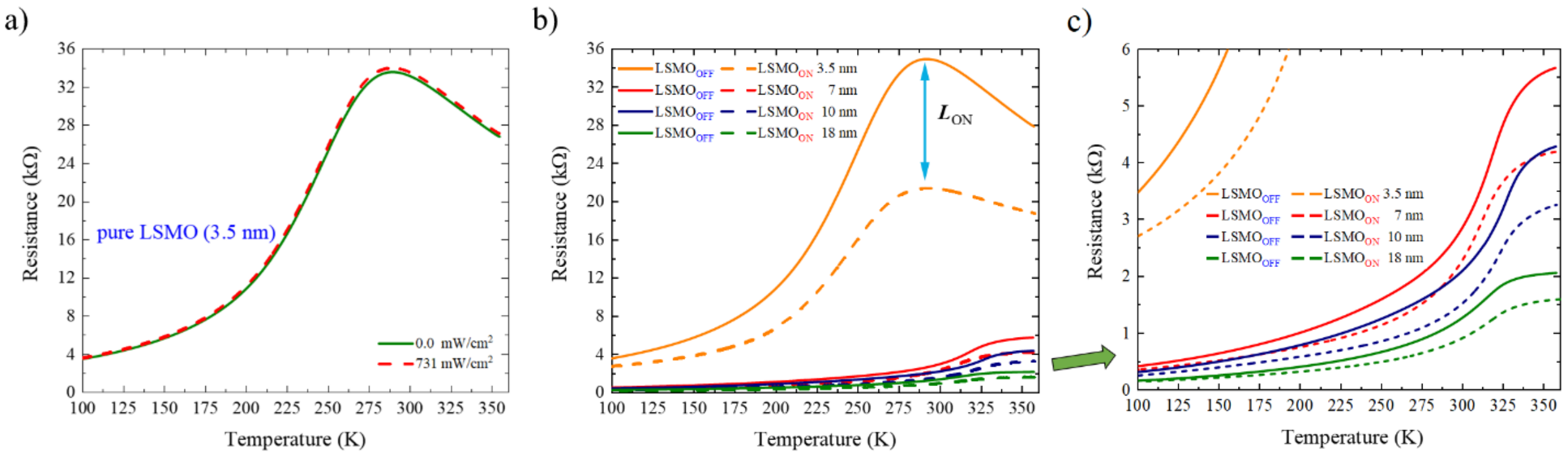

Figure S1: a) Electrical transport measurements as a function of temperature for pure LSMO with and without light. b) Electrical transport measurements as a function of temperature for CdS/LSMO heterostructures with different LSMO thicknesses with and without light. In both figures, the dashed lines correspond to the transport measurement with the maximum power density of 731 $mW/cm^2$. c) Electrical transport measurements as a function of temperature with a different axis scale enlarged from b) for better comparison and visibility.

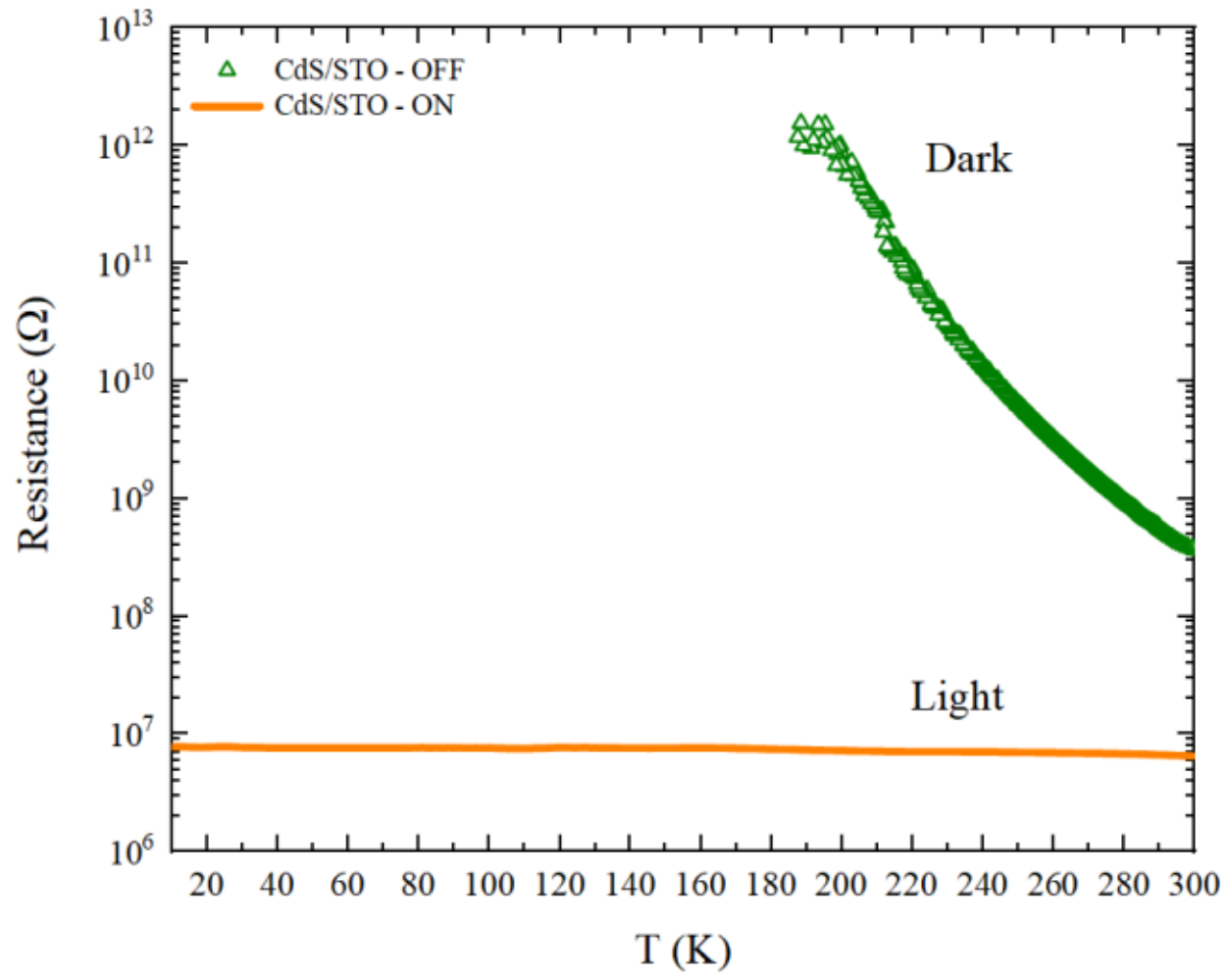

Figure S2: a) Electrical transport measurements as a function of temperature for pure CdS with and without light.

Figure S3 presents the morphological and structural characterization of pure LSMO samples. Morphological measurements were obtained from Scanning Probe Microscopy (SPM) at room temperature of the 3.5 nm LSMO exhibiting a flat surface with well-defined step-terrace features, with a roughness of 0.6 nm in a 5 x 5 $\mu m^2$ image. Figure S3(b) shows a θ–2θ scan of XRD taken from 3.5, 7, 10, and 18 nm LSMO films. The XRD confirms the growth of *c*-axis-oriented LSMO film with a slight lattice mismatch to the STO substrate. On the right side, the reflectivity of pure LSMO is presented with well-defined oscillations. The profile fitting of the low-angle X-ray measurements was made with the *Parrat32* code, which presents precise thicknesses with a roughness of less than 1 nm in each layer (see inset).

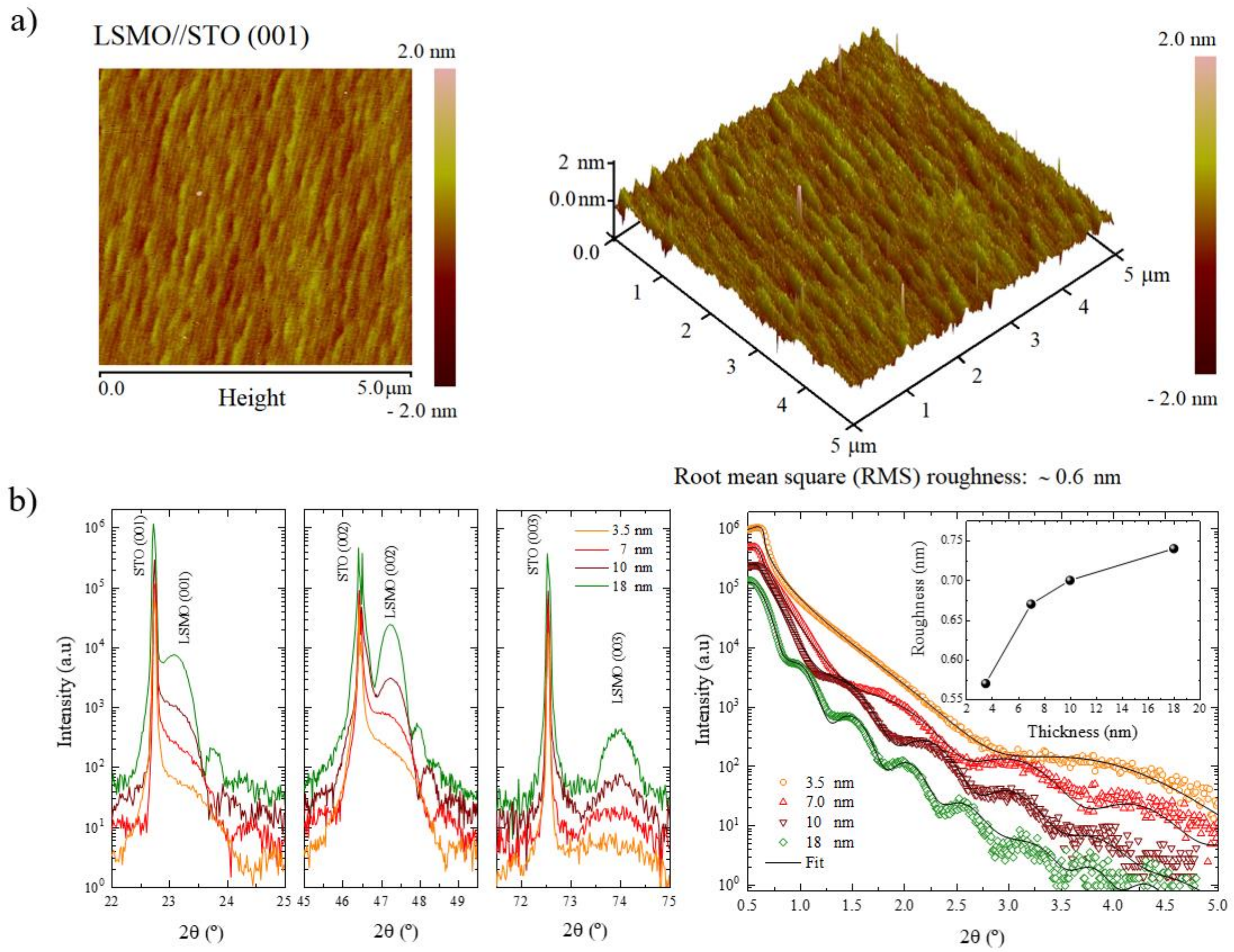

Figure S3: a) (Left) Topographical image of pure LSMO (3.5 nm) growth over $SrTiO_3$ (001) substrate, and 3D image of 5 x 5 $\mu m^2$ with a roughness of 0.6 nm (right). b) (Left) X-ray diffractogram (logarithmic intensity scale) of the pure LSMO (3.5, 7, 10, 18 nm). (Right) Reflectivity of the LSMO (3.5, 7, 10, 18 nm), the black lines correspond to the fit. The inset shows the evolution of the roughness when increasing the LSMO thickness.

Figure S4 presents the morphological and structural characterization of the CdS/LSMO sample with 3.5 nm LSMO, which has a roughness value of 5.6 nm in a 5 x 5 $\mu m^2$ image. Figure S4b) shows a θ–2θ scan of XRD taken from a 3.5 nm LSMO film with 80 nm CdS. The XRD confirms the growth of *c*-axis-oriented LSMO film and the characteristic peak of CdS (002). On the right is the reflectivity of CdS/LSMO measurements with the corresponding fitting (red line).

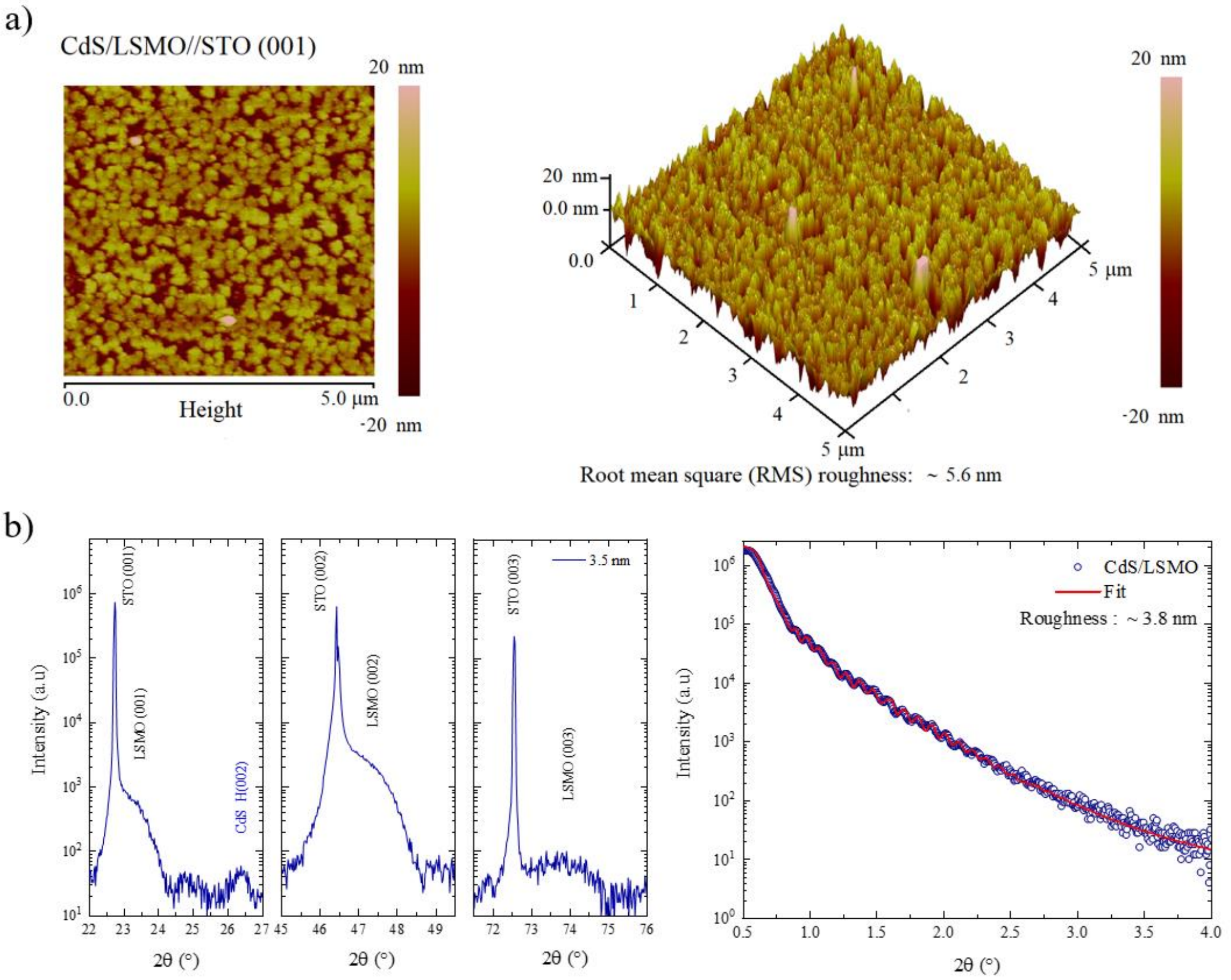

Figure S4: a) (Left) Topographical image of the hybrid CdS/LSMO (3.5 nm) growth over SrTiO3 (001) substrate, and 3D image of 5 x 5 μm$^2$ with a roughness of 5.6 nm (right). b) (Left) X-ray diffractogram (logarithmic intensity scale) of the pure LSMO (3.5 nm) with a CdS (002) plane at 26.5°. (Right) Reflectivity of the CdS (80 nm)/LSMO (3.5 nm), the red lines correspond to the fit with a roughness of 3.8 nm.

Figure S5 presents the normalized magnetization of the hybrid CdS/LSMO samples with LSMO (3.5, 7, 10, 18 nm). Note that the magnetization data for the ultra-thin sample is obtained from the magnetic field applied perpendicularly $\boldsymbol{H}_{\perp}$ to the film surface (the orange curve in Figure S5a)). The magnetic transition temperatures $T_c$ presented in Figure S5b) are obtained from the dM/dT magnetization of Figure S4a).

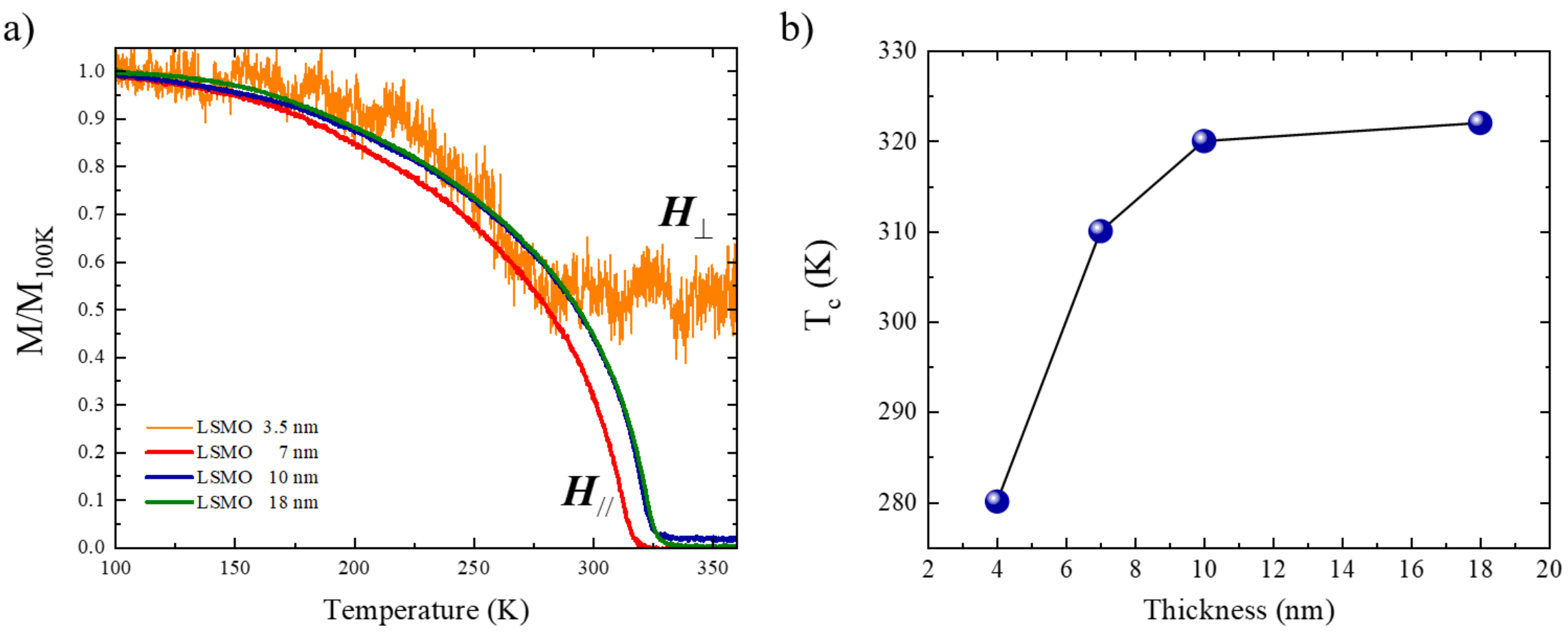

Figure S5: a) Normalized magnetization of the hybrid CdS/LSMO for different thicknesses of LSMO. *H* corresponds to the magnetic field applied parallel or perpendicular to the sample. b) Evolution of the $T_c$ obtained from the magnetization of the hybrid CdS/LSMO as a function of the thickness of the LSMO.

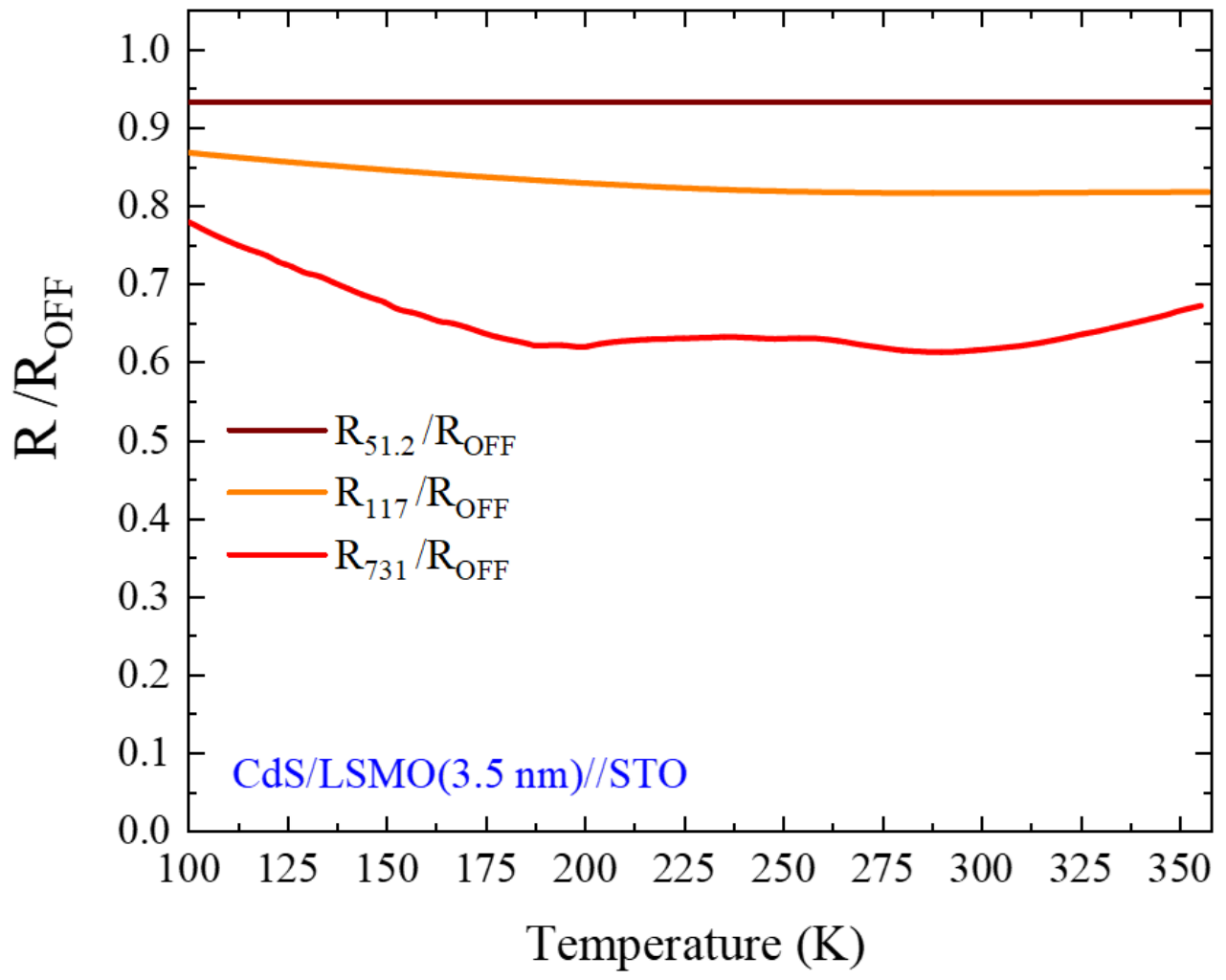

Figure S6: Electrical transport measurements normalized by $R_{OFF}$ as a function of temperature. Red curve represents the resistance with 731 mW/cm$^2$ illumination power density, Orange curve at 117 mW/cm$^2$ and Brown curve at 51.2 mW/cm$^2$. $R_{OFF}$ represent the curve with no light.

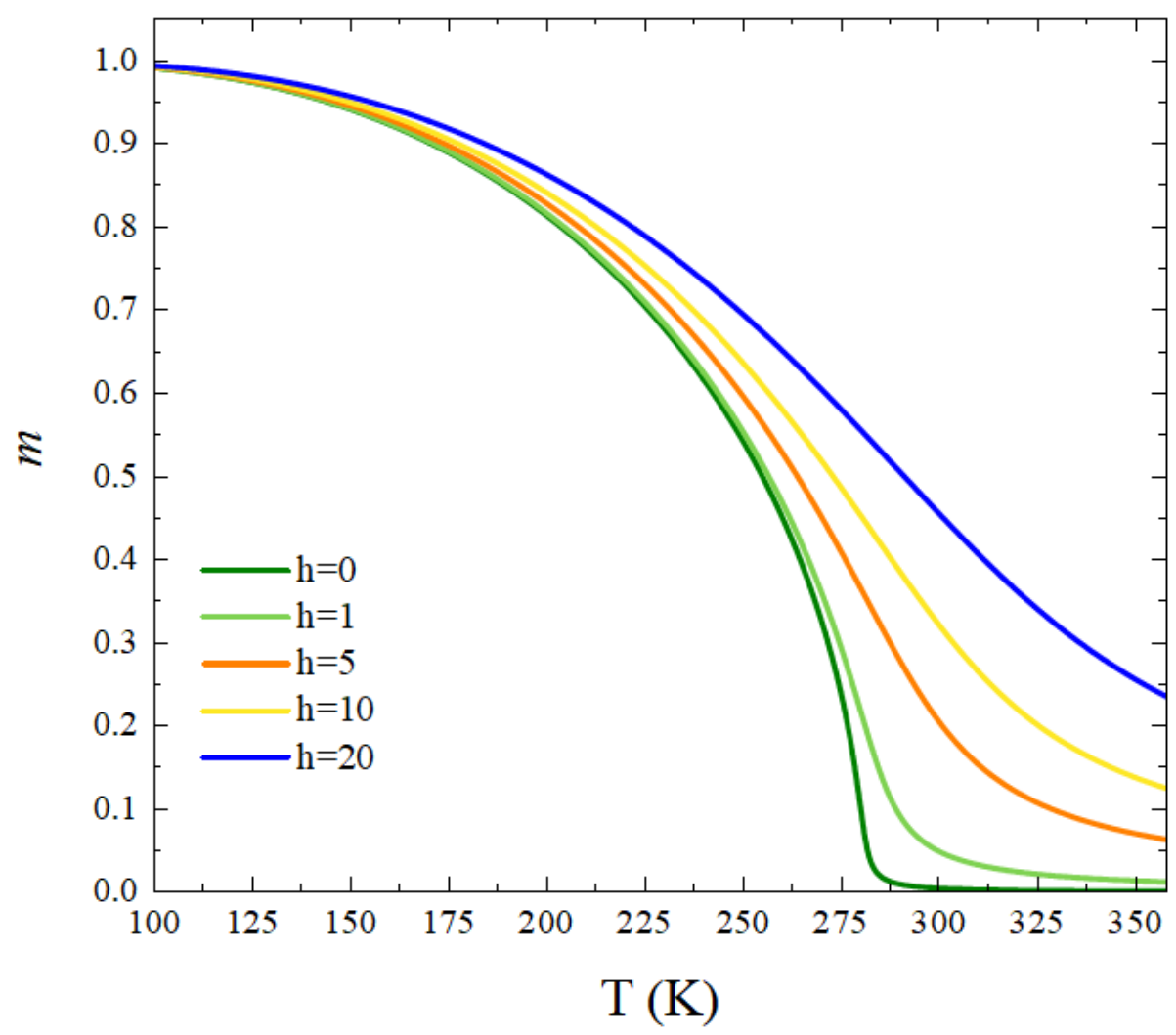

Figure S7: Calculations of change in the mean-field magnetization as a function of temperature for different magnetic fields *h*.

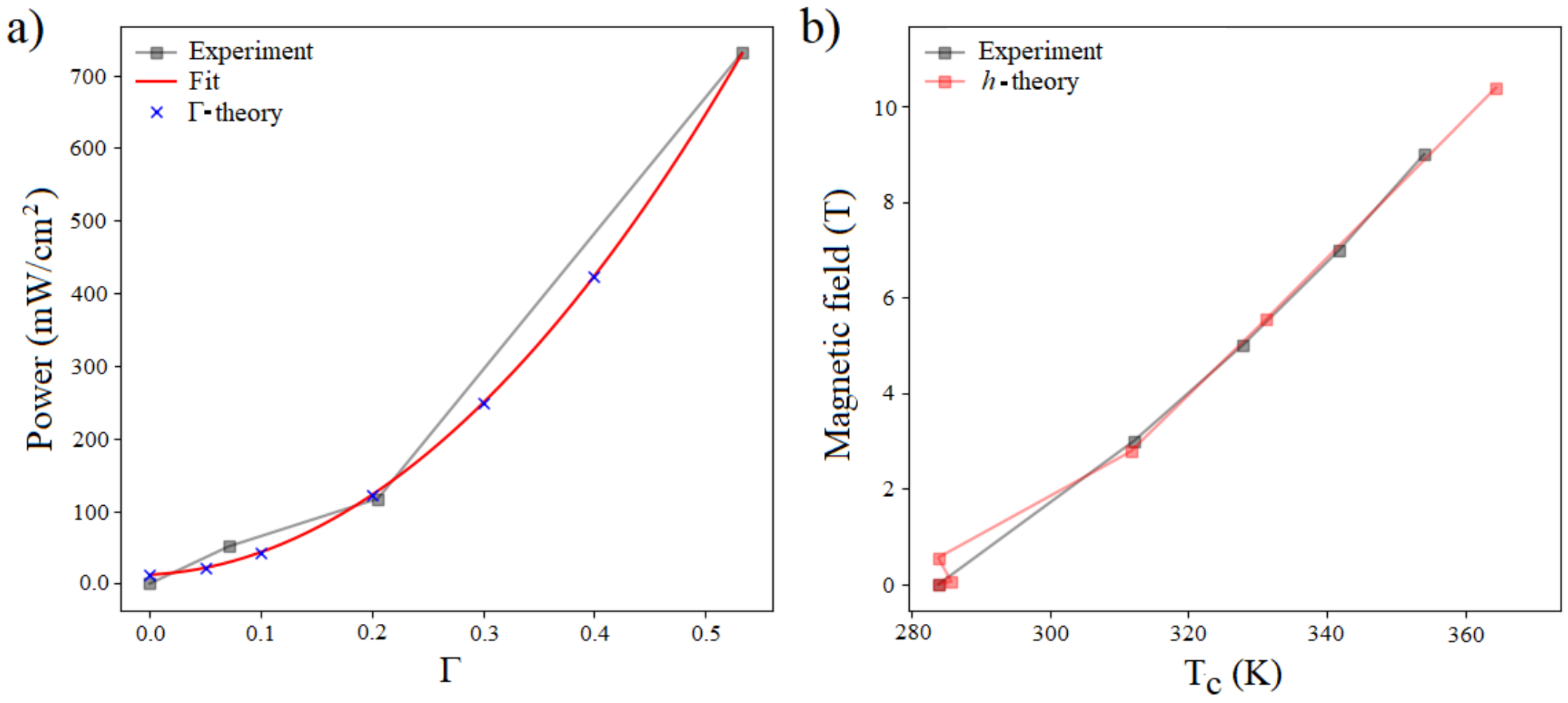

Figure S8: Comparison between experimental setup values and the theory. a) Light density power as a function of Γ. b) Magnetic field (experimental and theory) as a function of $T_c$.

This phenomenological scenario would be further understood by considering the physical properties of $La_{1-x}Sr_xMnO_3$ manganite, which is strongly dependent on hole doping produced by the chemical substitution of large divalent ions by small trivalent ion[1]. Below transition temperature, the $La_{0.7}Sr_{0.3}MnO_3$ exhibit a spin-gap opening revealing a half-metal behavior. As shown in Figure S9, the partially occupied spin-up band exhibits a metallic behavior. In contrast, the gap between occupied O(2p) and unoccupied Mn(3d) spin-down gap leads to an insulating spin-down band. The photodoping increases the charge carriers on the conducting band at the Fermi level, increasing the conductivity of the LSMO, but does not change the spin-gap of the insulating band; hence the critical temperature is preserved.

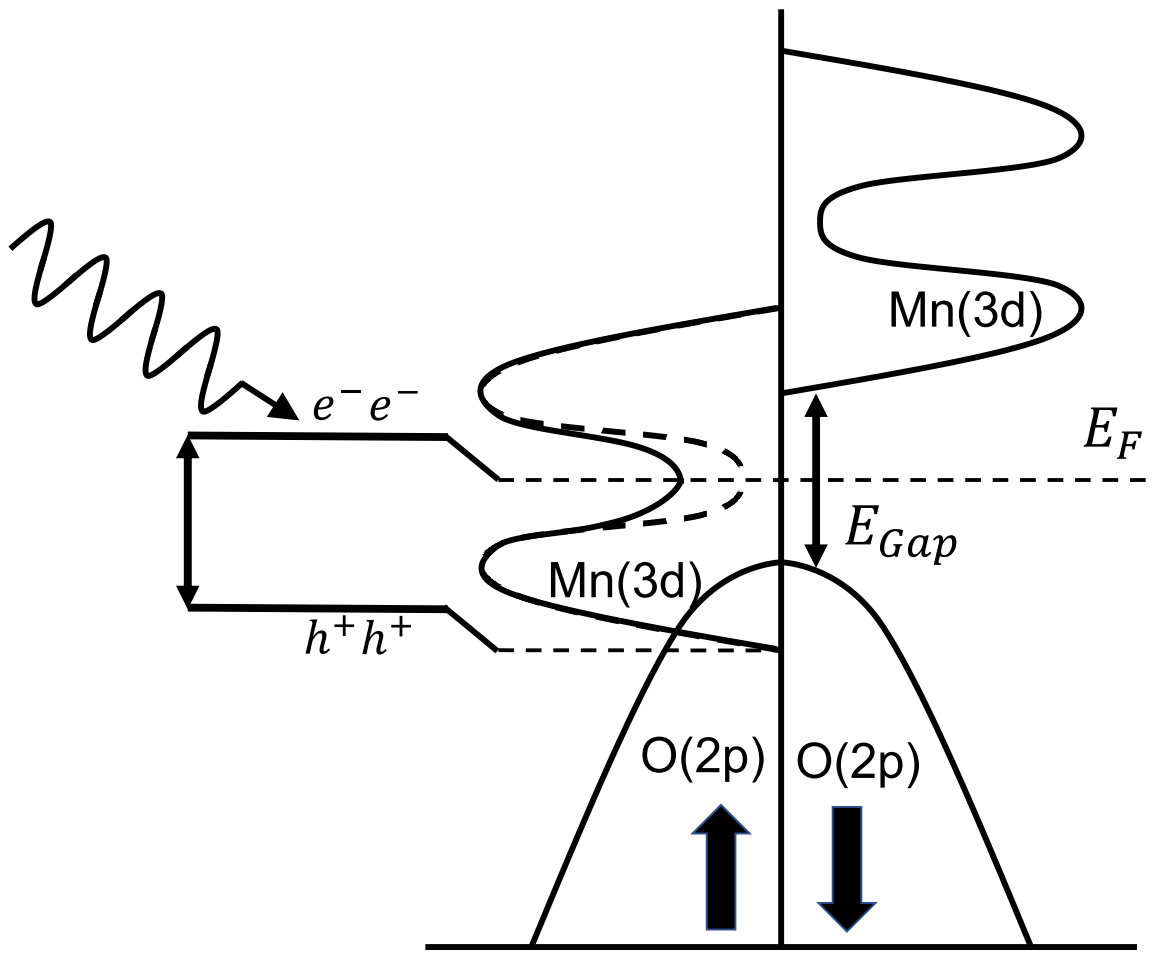

Figure S9: Schematic representation of energy diagram of LSMO half-metal configuration. Injection of photocarriers increases the density of states at the Fermi level favoring the electron hopping but preserving the spin gap. As a result, the Drude peak decreases as a function of light intensity at the same critical temperature.